# Photo-physics and electronic structure of lateral graphene/MoS$_2$ and metal/MoS$_2$ junctions

Shruti Subramanian[1,2], Quinn T. Campbell[1,3], Simon Moser[4,5], Jonas Kiemle[6], Philipp Zimmermann[6], Paul Seifert[6,7], Florian Sigger[6], Deeksha Sharma[8], Hala Al-Sadeg[1], Michael Labella III[9], Dacen Waters[10], Randall M. Feenstra[10], Roland J. Koch[4], Chris Jozwiak[4], Aaron Bostwick[4], Eli Rotenberg[4], Ismaila Dabo[1], Alexander Holleitner[6], Thomas E. Beechem[11], Ursula Wurstbauer[6,12], Joshua A. Robinson[1,2,13,14]

1. Department of Materials Science and Engineering, The Pennsylvania State University, University Park, Pennsylvania 16802, United States of America
2. Center for 2-Dimensional and Layered Materials, The Pennsylvania State University, University Park, Pennsylvania 16802, United States of America
3. Center for Computing Research, Sandia National Laboratories, Albuquerque, New Mexico 87185, United States of America
4. Advanced Light Source, E. O. Lawrence Berkeley National Laboratory, Berkeley, California 94720, United States of America
5. Physikalisches Institut and Würzburg-Dresden Cluster of Excellence ct.qmat, Universität Würzburg, 97074 Würzburg, Germany
6. Walter Schottky Institut and Physik Department, Technische Universität München, Am Coulombwall 4, 85748 Garching, Germany
7. ICFO - Institut de Ciencies Fotoniques, The Barcelona Institute of Science and Technology, Castelldefels, Barcelona, 08860, Spain
8. Department of Mechanical Engineering, The Pennsylvania State University, University Park, Pennsylvania 16802, United States of America
9. Nanofabrication Facility, The Pennsylvania State University, University Park, Pennsylvania 16802, United States of America
10. Department of Physics, Carnegie Mellon University, Pittsburgh, PA 15213, United States of America
11. Center for Integrated Nanotechnologies, Sandia National Laboratories, Albuquerque, New Mexico 87185, United States of America
12. Institute of Physics, University of Munster, 48149 Münster, Germany
13. 2-Dimensional Crystal Consortium, The Pennsylvania State University, University Park, PA 16802, United States of America
14. Center for Atomically Thin Multifunctional Coatings, The Pennsylvania State University, University Park, PA 16802, United States of America

*Abstract*

**Integration of semiconducting transition metal dichalcogenides (TMDs) into functional optoelectronic circuitries requires an understanding of the charge transfer across the interface between the TMD and the contacting material. Here, we use spatially resolved photocurrent microscopy to demonstrate electronic uniformity at the epitaxial graphene/molybdenum disulfide (EG/MoS$_2$) interface. A 10× larger photocurrent is extracted at the EG/MoS$_2$ interface when compared to metal (Ti/Au) /MoS$_2$ interface. This is supported by semi-local density-functional theory (DFT), which predicts the Schottky barrier at the EG/MoS$_2$ interface to be ~2× lower than Ti/MoS$_2$. We provide a direct visualization of a 2D material Schottky barrier through combination of angle resolved photoemission spectroscopy with spatial resolution selected to be ~300 nm (nano-ARPES)**

and DFT calculations. A bending of ~500 meV over a length scale of ~2-3 μm in the valence band maximum of $MoS_2$ is observed *via* nano-ARPES. We explicate a correlation between experimental demonstration and theoretical predictions of barriers at graphene/TMD interfaces. Spatially resolved photocurrent mapping allows for directly visualizing the uniformity of built-in electric fields at heterostructure interfaces, providing a guide for microscopic engineering of charge transport across heterointerfaces. This simple probe-based technique also speaks directly to the 2D synthesis community to elucidate electronic uniformity at domain boundaries alongside morphological uniformity over large areas.

Two-dimensional (2D) van der Waals semiconducting materials, such as transition metal dichalcogenides (TMDs) are of interest due to their strong light-matter interaction dominated by exciton phenomena.[1,2] Furthermore, van der Waals crystals are a highly tunable and versatile material systems that are poised to impact flexible and transparent electronics,[3–5] optoelectronics,[6–9] spin-and valleytronics,[10] catalysis,[11–13] biomedical[14–16] and sensor[17,18] applications, and next-generation quantum materials.[19–23] Additionally, semiconducting 2D crystals can be engineered by external stimuli such as doping,[24,25] strain,[26,27] defects,[28] pressure,[29] environment,[30] by interaction with light[31] or in moiré superlattice structures.[32] Key to the integration of semiconducting TMDs into functional circuitries is understanding the charge transfer across a 2D semiconductor and metallic heterojunction allowing for them to be precisely engineered to the specific functionality. In this regard, lateral junctions between molybdenum disulfide ($MoS_2$) and conventional metals, as well as $MoS_2$ and graphene are utilized in electronic and optoelectronic device schemes with graphene, providing better contact behavior[33,34] and tunability due to its semi-metallic behavior.[35]

Here, we investigate the local charge transfer characteristics and the lateral heterogeneity of metal/$MoS_2$ and graphene/$MoS_2$ junctions *via* scanning photocurrent microscopy, and elucidate the band bending of both structures by a combination of angle resolved photoemission spectroscopy with spatial resolution selected to be ~300 nm (nano-ARPES) and DFT calculations. Illustrating the close correlation between experimental demonstrations and theoretical predictions of barriers at graphene/TMD interfaces allows us to develop a comprehensive understanding of charge transfer, aiding the microscopic engineering of charge transport across heterointerfaces. Direct visualization of built-in fields over large areas at heterostructure interfaces *via* scanning photocurrent microscopy speaks directly to the 2D synthesis community to correlate and demonstrate electronic uniformity alongside morphological uniformity.

*Results/Discussion*

Efficient electron-hole pair ($e^-$-$h^+$) generation in $MoS_2$-based devices upon photoexcitation is responsible for a single-layer $MoS_2$-based photodetector exhibiting a photoresponsivity reaching 880 A/W.[36] Despite the rich photo-physics, performance of $MoS_2$ devices that rely on efficient photocurrent generation are limited by large exciton binding energies and resulting excited state decay.[37–40] Slow photoresponse dynamics observed in $MoS_2$-based devices necessitates a comprehensive understanding of photocurrent dynamics and charge carrier recombination in MoS$_2$.[36,41] High performance optoelectronic detectors require efficient separation of charge carriers following photogeneration to prevent them from recombination before being collected at the electrodes. This separation of e$^-$-h$^+$ pairs is facilitated by the existence of a built-in electric field in the system, which leads to the idea of constructing an electronic barrier in the system, away from the metal contacts, to aid the separation of charge carriers. In this work, we examine an epitaxial graphene (EG)/MoS$_2$ lateral heterostructure exhibiting a uniform built-in electric field at the interface while using 2-3 layer EG as the contact material and allowing for an efficient separation of photogenerated charge carriers. EG/MoS$_2$ lateral heterostructures are synthesized in the same controllable manner as previously demonstrated by Subramanian *et al.* with the resulting structure having a ~50-200 nm overlap of the multi-layer MoS$_2$ onto the edge of the patterned EG[33] as shown in the schematic of Figure 1(a) and 1(b).

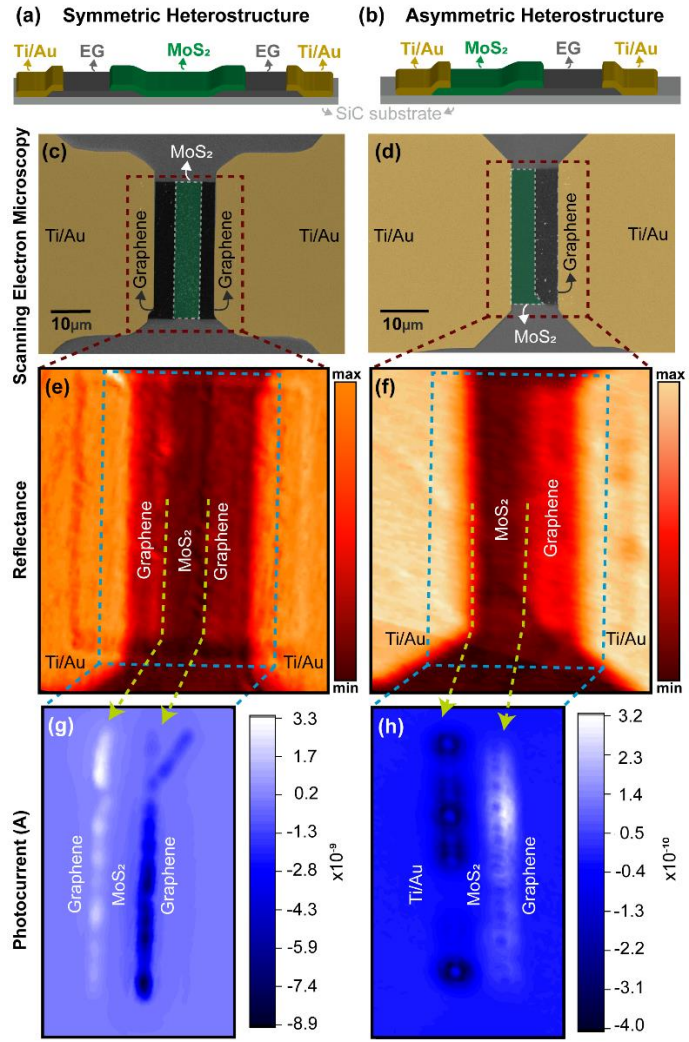

Spatial photocurrent mapping reveals the intrinsic photoresponse at the EG/MoS$_2$ and the metal/MoS$_2$ interfaces. The built-in electric field and photogenerated carrier extraction are compared utilizing photocurrent mapping of two synthesized heterostructure systems (Figure 1): (i) A symmetric lateral heterostructure consisting of metal/EG/MoS$_2$/EG/metal as shown in Figure 1(a,c), and (ii) An asymmetric lateral heterostructure consisting of metal/MoS$_2$/EG/metal as shown in Figure 1(b,d). The spatial resolution of photocurrent in 2D heterostructures allows for discerning local variations in the electric field. An optical probe maps the reflection from the two heterostructure systems, as shown in Figure 1(e) and Figure 1(f) for the symmetric and asymmetric heterostructure systems, respectively, and this is directly coupled with the photocurrent maps. The different components of the heterostructure devices have different optical contrasts as seen in the reflectance maps. Photocurrent is observed at the heterostructure interfaces as shown in Figure 1(g) and 1(h), and is attributed to two mechanisms contributing to the photocurrent as extensively studied by Parzinger *et al.* – the primary contributor being the photovoltaic effect, *i.e.*, the generation of e$^-$-h$^+$ pairs and their subsequent spatial dissociation due to an electric field, and the secondary contributor being the photo-thermoelectric effect that allows for electrons to have a flux in response to the established temperature gradient by laser irradiation.[9]

**Figure 1:** (a) Schematic showing the synthesized symmetric EG/MoS$_2$/EG device lateral heterostructure with the overlap of the MoS$_2$ onto the edge of the patterned EG. (b) Schematic of the asymmetric heterostructure device with metal/MoS$_2$/EG configuration. (c) False-colored scanning electron micrograph (SEM) of the symmetric heterostructure with EG on both sides of the MoS$_2$. (d) SEM of the asymmetric heterostructure with EG on one side and Ti/Au metal on the other. (e) Reflectance map of the symmetric heterostructure zoomed in from (c). (f) Reflectance map of the asymmetric heterostructure zoomed in from (d). (g) Photocurrent map (at zero applied bias) of the EG/MoS$_2$/EG heterostructure displaying photo-activity at the EG/MoS$_2$ interface. No photoactivity is seen at the metal/EG interface. Dotted lines are drawn to guide the eye to the photoactive interface. (h) Photocurrent map (at zero applied bias) of the metal/MoS$_2$/EG heterostructure displaying "patchy" photo-activity at the metal/MoS$_2$ interface as compared to the EG/MoS$_2$ interface. 10× reduction in overall photocurrent is observed when compared to (g).

The presence of a depletion region at the heterostructure interface leads to a measurable photocurrent upon illumination. For MoS$_2$ multilayers, if the excitation energy is less than the direct bandgap but more than the indirect bandgap, e$^-$-h$^+$ pairs will still be generated, but with reduced probability because this transition requires a phonon to change momentum.[42,43] The lateral heterostructures studied here are dominated by multilayer MoS$_2$, especially at the EG/MoS$_2$ heterostructure interface.[33] We illuminate the heterostructures using the emission line of an Ar/Kr laser at 488 nm (2.54 eV), whose energy is always greater than the direct bandgap of MoS$_2$.

The time-integrated photocurrent at heterostructure interfaces is primarily attributed to photovoltaic effect.[9] Apart from an observation of a measurable photocurrent at the heterostructure interfaces, there are further subtleties visible from the photocurrent maps in Figure 1. Based on Figure 1(g), the photo-active interface is the EG/MoS$_2$ interface, indicating the presence of a Schottky barrier that separates photogenerated charge carriers.[44] Additionally, the metal/EG interface in the heterostructure system is not photo-active. This is contrary to the observations of photocurrent in exfoliated and CVD graphene at metal contacts where the Schottky barrier is formed.[45–48] Here, the intrinsic polarization of the silicon carbide (SiC) substrate electrostatically dopes EG *n*-type, reducing the Schottky barrier at the metal(Ti/Au)/EG interface. In contrast, both exfoliated and CVD graphene are typically *p*-type,[49,50] forming a much larger Schottky barrier with the Ti/Au metal stack, thus enabling significant space-charge regions at the interface. Epitaxial graphene on SiC also exhibits less surface impurities compared to CVD and exfoliated graphene since it does not undergo polymeric transfer processes, which may be a contributing factor to the reduced Schottky barrier at the metal (Ti/Au)/EG interface. All measurements are performed in vacuum to avoid the environmental doping of the 2D materials leading to changes in their Schottky barriers. The photocurrent landscape in the symmetric heterostructure is uniform through the length of the device as compared to the asymmetric heterostructure (See supplementary section S.1). Importantly, a uniform photocurrent exists along the length of the EG/MoS$_2$ interface regardless of device symmetry (Figure 1(g,h)), noting the existence of a uniform built-in electric field due to a pristine EG/MoS$_2$ interface. In contrast, the metal/MoS$_2$ interface exhibits a non-uniform photocurrent signal, (Figure 1(h)) (Photocurrent standard deviation ($\sigma$) for the metal/MoS$_2$ > 1.5× EG/MoS$_2$ - See supplementary section S.1), possibly resulting in current crowding and localized heating of the device. We note that the laser illumination occurs through the contact layer (EG or 20 nm of Ti/Au – 5 nm Ti + 15 nm Au). A non-negligible part of the light, ~ 25%, is absorbed by the Ti/Au metal contact stack (Calculations done on www.filmetrics.com/reflectance-calculator based on the complex-matrix form of the Fresnel equations).[51–55] This can lead to heating of the metal contact and may provide additional contributions to the observed photocurrent *via* thermal effects that could have opposite sign and spatial extension across the junction.[9] Together with the reduced dark current, these additional contributions may be responsible for the reduced photocurrent signal in the asymmetric heterostructure when compared to the symmetric heterostructure. The improved uniformity of the built-in field at the EG/MoS$_2$ interface (Figure 1(d)) allows for efficient charge separation and an improved electronic conduction through the entire heterostructure system, as evidenced by >10× increase in the measured photocurrent compared to the asymmetric device (See supplementary section S.1). This trend is observed in all tested devices, spanning different samples, and directly corroborates our previous work where EG has been shown to be a better contact to semiconducting MoS$_2$ compared to conventional metals.[33]

Semi-local density-functional theory (DFT) predicts the Schottky barrier at the EG/MoS$_2$ interface to be ~2× lower than Ti/MoS$_2$. Prediction of the Schottky barrier from first-principles necessitates, initially, calculations of work functions ($\phi$) of the individual components – pristine graphene ($\phi_{gr}$), MoS$_2$ ($\phi_{MoS_2}$), and Ti ($\phi_{Ti}$) – and the relaxed heterointerfaces ($\phi_{EG/MoS_2}$, $\phi_{Ti/MoS_2}$), summarized in Table 1. In this case, however, the work function of epitaxial graphene on SiC substrate, referred to as EG ($\phi_{EG}$) is dominated by the intrinsic polarization in the SiC substrate, necessitating consideration of the SiC work function ($\phi_{SiC}$). Previous first-principles calculations suggest work functions of undoped graphene, MoS$_2$, and Ti to be 4.23 eV, 4.05 eV, and 4.38 eV, respectively.[56–58] Here the trend is similar, but the absolute value is consistently lower by ~0.3 eV due to different parameterizations of the exchange-correlation functionals and van der Waals corrections in other works. Since the choice of parameterizations affect the absolute values of the work function but not the relative differences between them, the barrier calculations are considered accurate. $\phi_{gr}$ and $\phi_{SiC}$ differ by ~0.15 eV, leading to an equilibrium $\phi_{EG}$ that is calculated from the SiC/EG interfacial dipole as 3.99 eV.[59]

**Table 1:** The work functions of both isolated materials and interfaces as calculated from DFT. A comparison with alternate DFT functionals is included in Supplementary S.2.

| Material | Work function (eV) |
|---|---|
| Pristine graphene ($\phi_{gr}$) | 3.92 |
| MoS$_2$ ($\phi_{MoS_2}$) | 3.78 |
| SiC ($\phi_{SiC}$) | 4.06 |
| Ti ($\phi_{Ti}$) | 4.02 |
| EG on SiC ($\phi_{EG}$) | 3.99 |
| EG/MoS$_2$ interface ($\phi_{EG/MoS_2}$) | 3.55 |
| Ti/MoS$_2$ interface ($\phi_{Ti/MoS_2}$) | 3.21 |

The total voltage drop $\Delta\phi$ across the junction is defined as $\Delta\phi = \phi_{EG} - \phi_{EG/MoS_2}$, leading to $\Delta\phi = 0.45$ V and 0.81 V for EG/MoS$_2$ and Ti/MoS$_2$, respectively. To calculate the height of the Schottky barrier ($\phi_B$), we apply a technique developed for semiconductor interfaces, known to take into account Fermi level pinning[60,61] (See supplementary section S.2):

$$\phi_B = \frac{\varepsilon\varepsilon_0}{2e_0 N_d} \left|\frac{d\varphi}{dz}(z_c)\right|^2 \qquad \text{(Equation 1)}$$

where $\varphi$ is the electric potential of the equilibrium interface along the $z$-axis perpendicular to the plane of the interface, $e_0$ is the electron charge, $N_d$ is the three-dimensional defect density in the semiconductor, $\varepsilon$ and $\varepsilon_0$ are the dielectric constant of MoS$_2$ ($\varepsilon = 4.3$)[62] and the vacuum permittivity, respectively. $z_c$ is the location along the transverse axis marking the transition between the quantum mechanical region of the interface model and the one-dimensional continuum (Mott-Schottky) description of the bulk semiconductor; $z_c$ is located two layers within the semiconductor to fully capture the effects of the interface before transitioning to a bulk model. The only unknown in equation 1 is $N_d$. Following the method of Kim *et al.*,[63] we use the density that produces the best fit with the experimental nano-ARPES shown later, $N_d = 4 \times 10^{12}$ cm$^{-3}$.

The Schottky barrier of the EG/MoS$_2$ interface is predicted to be 0.44 eV. This compares favorably with the previous predictions (0.4 eV) for undoped graphene/MoS$_2$ junction,[64] and reasonably with the prediction of 0.6 eV by Jin *et al.*, using many-body perturbation (G$_0$W$_0$) calculations and a single monolayer of both graphene and MoS$_2$.[65] A notable feature of the EG/MoS$_2$ interface in this study is the high fraction of $\Delta\phi$ compensated by the Schottky barrier at the interface. One explanation is that atomically "clean" interfaces, such as the EG/MoS$_2$, have less covalent and ionic bonds and thus lower surface states, causing a larger fraction of potential offset $\Delta\phi$ to be taken up by the Schottky barrier $\phi_B$. In contrast, the Schottky barrier of the Ti/MoS$_2$ interface is predicted to be 0.79 eV. The voltage drop due to surface states at the interface barrier is ~2× larger for Ti/MoS$_2$ than for EG/MoS$_2$, due to increased charge trapping at the interface, indicating a rougher interfacial contact with more ionic and covalent bonds. The ionic and covalent nature of the interfaces was confirmed with Bader analysis, see supplementary section S.2 for details.

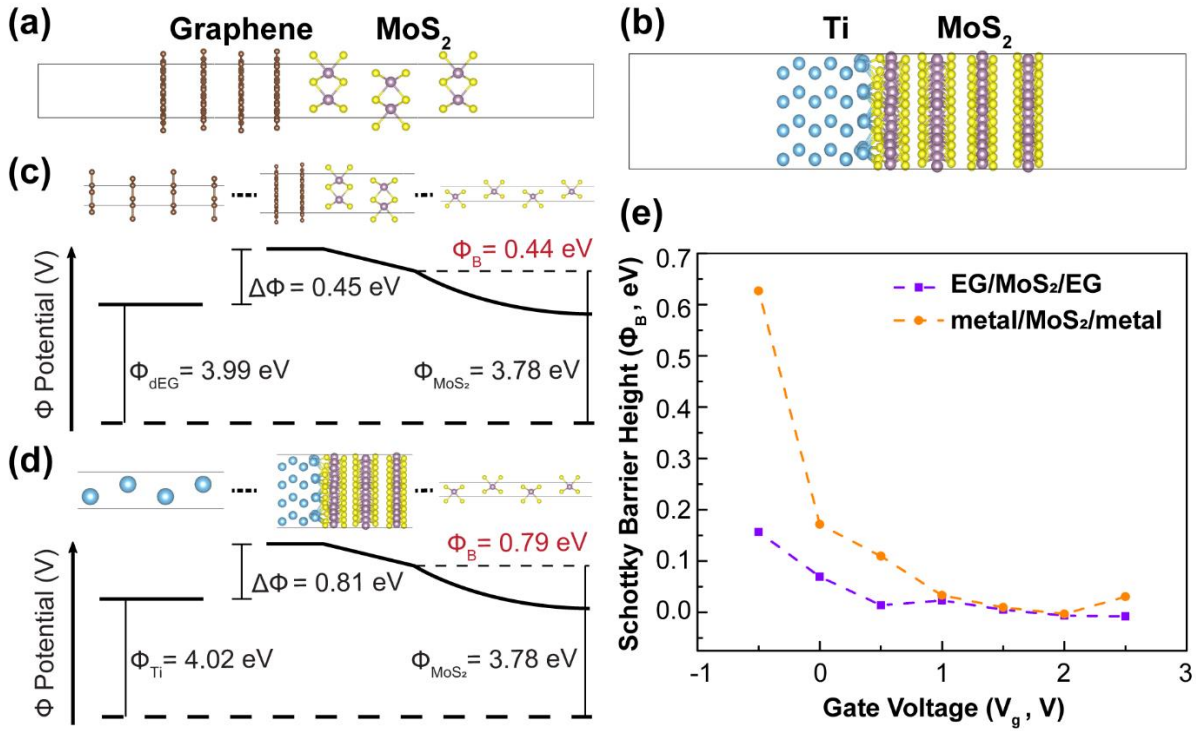

**Figure 2:** Structures of the (a) EG/MoS$_2$ interface and (b) Ti/MoS$_2$ interface from electronic-structure DFT calculations. (c) Diagram of the potential across the EG/MoS$_2$ heterointerface. The interfacial potential difference $\Delta\phi$ is compensated by a potential drop at the interface (calculated from semi-local DFT) and by a Schottky barrier $\phi_B$ (calculated using DFT-continuum embedding techniques; see Ref. 60, 61). The equilibrium voltage distribution is determined by self-consistently matching the final work functions of the doped graphene and the interface. (d) The same diagram as in panel (c), showing a higher value of predicted Schottky barrier at the Ti/MoS$_2$ interface. (e)

Experimental Schottky barriers extracted using Arrhenius plots from temperature-dependent gated current measurements (*I-V-T*) show a lower value of the Schottky barrier at the EG/MoS$_2$ interface as compared to the Ti/MoS$_2$ interface.

Temperature dependent gated current (I-V-T) measurements validate Schottky barrier predictions for EG contacted MoS$_2$ and Ti/Au metal contacted MoS$_2$. Electrostatic double layer (EDL) gating is implemented in a helium-cooled cryostat and the temperature range used is 3-300K and the gate voltage ($V_g$) range used is -2.5V to +2.5V. The Arrhenius equation is used to extract the apparent Schottky barrier heights for the two structures (See supplementary section S.3 for further details). EG/MoS$_2$ heterostructure interface is shown to have a reduced Schottky barrier as compared to Ti/Au contacted MoS$_2$ in Figure 2(e) – this result corroborates the photocurrent measurements and the theory calculations of Schottky barrier heights. The difference in the experimental and predicted values is due to the complex interaction of the non-ideal materials with their substrates and dielectric environments. For instance, while we do see significant covalent bonding and intermixing at the Ti/MoS$_2$ interface in our calculations, previous literature has found a large interlayer of Ti and MoS$_2$ which is computationally prohibitive to predict.[66] This may lead to an increase in the predicted Fermi level pinning at this interface and a decrease in the predicted Schottky barrier. Having established that the EG/MoS$_2$ heterointerface is more efficient for charge separation *via* spatially resolved photocurrent measurements and Schottky barrier extraction, it is essential to understand the specific movement of charge carriers at the EG/MoS$_2$ heterointerface.

Angle resolved photoemission spectroscopy with selected spatial resolution of ~300 nm (nano-ARPES) visualizes spatial variation in electronic band structures at the EG/MoS$_2$ heterostructure interface to understand the photo-generated charge transfer. A lower Schottky barrier at the photo-active EG/MoS$_2$ interface is coupled with a uniform built-in electric field aiding efficient charge separation. The spatial variations in the electronic band structure across the lateral EG/MoS$_2$/EG interface (as shown in Figure 3(a)) are investigated at steady state by means of synchrotron-based nano-ARPES. The lateral heterostructure in this experiment is intentionally not allowed to coalesce through the width of the channel between the EG, to enable the study of variations in the electronic structure with edges, where the X-ray probe is focused by a Fresnel zone plate and scanned across the sample. The ARPES band structure I(E, *k*) is measured at every spatial point coordinate (x,y) creating a 4-dimensional (4D) dataset; E is the binding energy referenced to the Fermi level and *k* is the in-plane momentum along the orientation of the analyzer entrance slit and I is the intensity of the obtained signal. See supplementary section S.4 for the schematic of the nano-ARPES setup. In this scanning mode, nano-ARPES is particularly suited for flat conductive samples, providing a well-defined surface normal, and thus a well-preserved momentum resolution. Room temperature nano-ARPES results obtained with 98 eV photons and an X-ray spot size of ~300 nm FWHM across the MoS$_2$ channel are summarized in Figure 3.

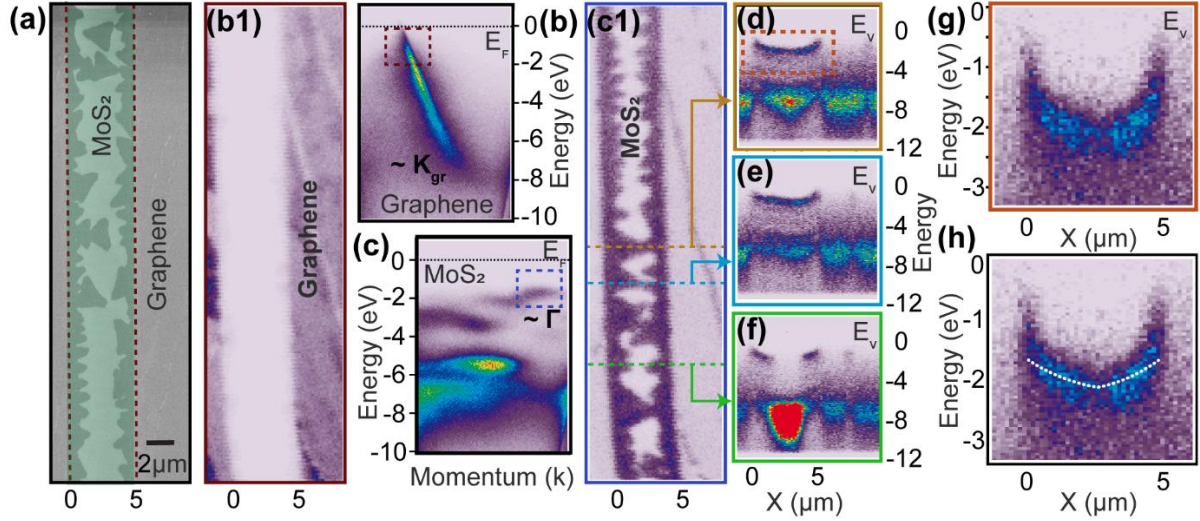

**Figure 3:** (a) False colored scanning electron micrograph of an intentionally non-coalesced EG/MoS$_2$ lateral heterostructure. The MoS$_2$ nucleates from the edges of the EG and grows outwards. Hence, the darker triangular features are the MoS$_2$ with the lighter region being the exposed SiC substrate. (b) and (c) show two E vs $k$ ARPES band structure maps focusing specifically on EG and MoS$_2$, respectively. (b1) and (c1) use the E vs $k$ ARPES band structure maps shown in (b) and (c) to obtain the spatial distribution of EG and MoS$_2$. Darker contrast corresponds to higher intensity. Upon overlaying (b1) and (c1), surface morphology similar to that in (a) is revealed. Since these measurements are made with separate instruments, we have found similar regions of the same sample but they are not exactly superimposable. Horizontal cuts at different y-values are taken for MoS$_2$ shown in (d), (e) and (f). (f) specifically focuses on an uncoalesced region to see the spatial variation of the bands. (g) is a zoomed in version of (d) displaying an ~500 meV voltage drop from the EG/MoS$_2$ interface to the center of the channel. (h) Band bending as predicted by first-principles calculations in white overlaid on the experimental nano ARPES results presented in Figure 3(g). The agreement indicates that the grown interface is relatively clean, leading to a low amount of surface states and a considerable Schottky barrier.

The valence band maximum in MoS$_2$ bends ~ 500 meV over a length scale of 2-3 µm from the edge of the EG/MoS$_2$ junction to the center of the channel. Figure 3(b) and 3(c) show two ARPES measured E vs $k$ band structure maps. A darker contrast indicates a higher intensity (scale bars for each panel is shown in supplementary section S.4). Figure 3(b) focuses on a sample region containing only graphene, showing the well characterized π-band that crosses the Fermi level at K. Using the 4D dataset to extract the total intensity from the (E,$k$) region that highlights this π-band (dark red dashed rectangle in Figure 3(b)), and plotting the result as a function of (x,y), the spatial distribution of graphene across the scanned sample area is obtained as shown in Figure 3(b1). Similarly, focusing on a region containing only MoS$_2$ (Figure 3(c)), and integrating over (E,$k$) such as to only highlight MoS$_2$ valence band states (blue rectangle in Figure 3(c)), we can reveal the distribution of MoS$_2$ across the sample as shown in Figure 3(c1). Upon overlaying, Figures 3(b1) and 3(c1) reproduce the surface morphology as obtained from a scanning electron micrograph on a similar channel in Figure 3(a). Since these measurements are made with separate instruments, we have found similar regions of the same sample but they are not exactly superimposable. The 5 µm channel etched in the EG is filled with triangles of MoS$_2$. An EG/MoS$_2$ heterostructure synthesized in this manner has an overlap of the MoS$_2$ on the EG at the patterned edge for a length scale of ~200 nm, on the order of the nano-ARPES resolution, and consequently is thicker at the patterned edge.[33] The intentionally non-coalesced lateral heterostructure system exhibits regions where the MoS$_2$ bridges the channel, which serve to explore band bending within the channel. Step edges on the SiC substrate are readily observed as lines with lighter contrast in Figure 3(b1). Corresponding dark lines in Figure 3(c1) shows the favorable nucleation of $MoS_2$ at the step edges of EG, which are known to be thicker and more defective compared to the EG on the terraces.[67] To understand how the electronic band structure changes at the EG/$MoS_2$ heterostructure interface, we focus on the $MoS_2$ valence band maximum (VBM) and observe how it bends upon moving away from the interface. Horizontal cuts at select y-values, with differing $MoS_2$ coverage, are taken from Figure 3(c1), marked by the yellow, cyan and green lines, allowing for E *versus* x plots (Figures 3(d), 3(e) and 3(f)) mapping the variation of the $MoS_2$ VBM across the EG/$MoS_2$/EG interface. Figure 3(g) is a close-up of Figure 3(d), where the variation of the VBM is seen along a cut (yellow line in Figure 3(c1)) where the channel is completely coalesced with $MoS_2$. Here, it is observed that the VBM bends strongly downwards, from approximately -1.5 eV at the interface to -2 eV near the channel center. The screening at this position is isotropic (Figure 3(g)), leading to symmetric band bending with respect to the center of the channel. The band bending in Figure 3(e) is similar in magnitude but highly asymmetric, possibly as a result of anisotropic screening due to variation in domain sizes and growth modes in both directions. Figure 3(f) shows the cut along the green line in Figure 3(c1), where the $MoS_2$ channel is not coalesced and the VBM is interrupted. The precise shape of the band bending is a function of the specific electrostatic environment, determined by coalescence in the channel as well as charge carrier screening from the EG and the SiC substrate. We assume a rigid band model and extrapolate the bending of the VBM observed *via* nano-ARPES to be the same as the bending of the conduction band minimum (CBM). The direction of band bending indicates that the electronic charges are transferred out from the $MoS_2$ into the EG upon formation of the heterostructure, indicating the $MoS_2$ donates electrons to the EG in order to attain equilibrium. This band bending is thus, a direct observation of the Schottky barrier at the EG/$MoS_2$ lateral heterostructure interface.

Density functional theory calculations can help interpret the spatial distribution of the potential at the EG/$MoS_2$ heterostructure interface. Following the Schottky barrier calculations (Figure 2), the potential $\varphi(x)$ within the semiconductor can be expressed as[68]:

$$\varphi(x) = -\frac{e_0 N_d}{2\varepsilon\varepsilon_0}x^2 - \frac{d\varphi}{dz}(z_c)x + \phi_s + \varphi_0 \qquad \text{(Equation 2)}$$

where $x$ is the distance from the EG/$MoS_2$ interface, and $\varphi_0$ is a constant to fulfill the boundary condition at the midpoint of the leftmost edge of the band region in Figure 3(g). The expected band bending is plotted over the experimental results in Figure 3(h). Since the $MoS_2$ is symmetrically terminated on both sides with an EG junction, the band bending in Figure 3(h) is shown according to Equation 2 until the midpoint of the $MoS_2$ region, and then the potential is symmetrically mirrored on the other side. In our calculations, with only ~2.5 µm of space on each side for the potential to decrease, the dopant density is not high enough to reach the theoretically predicted height of the Schottky barrier, instead only forming a barrier of 0.38 eV. This leads to some uncertainty in whether the overall potential should be shifted slightly down to match the Fermi level of the EG, but the range of uncertainty nevertheless is small compared to the extent of the observed band regions. The agreement of the theoretical potential with its nano-ARPES counterpart at the carrier concentration of 4 x $10^{12}$ $cm^{-3}$ strongly indicates that the synthesized interface is atomically clean with a low density of defects. This work highlights the capabilities of first-principles methods to understand behaviors at semiconductor junctions and facilitate the interpretation of experiments, providing physical insights and predictive trends.

*Conclusion*

EG/MoS$_2$/EG symmetric heterostructure has >10× larger photocurrent at the EG/MoS$_2$ interfaces, as compared to metal/MoS$_2$/EG asymmetric heterostructure, with a uniform built-in field through the length of the symmetric heterostructure device. The electronic Schottky barrier at the EG/MoS$_2$ interface is predicted to be ~2× lower than Ti/MoS$_2$ using DFT, corroborated by experimental I-V-T measurements. To further understand the transfer of photo-generated charge at the EG/MoS$_2$ interface, spatial variations of electronic bands are investigated using nano-ARPES. The valence band maximum in MoS$_2$ bends ~500 meV over a length scale of 2-3 µm, matching theoretical calculations. This comprehensive understanding of the photo-physics and optoelectronic properties of the EG/MoS$_2$ lateral heterostructure system can be extrapolated to other systems in order to build a library of photo-active heterostructure interfaces with properties tailored for specific optoelectronic applications. We have highlighted the utility of first-principles calculations to interpret the electrical response at heterostructure interfaces. The photocurrent measurements presented here are demonstrated to be a simple probe to measure the electronic uniformity of the synthesized lateral heterostructure interfaces. This simple probe-based technique can be adopted easily by the 2D materials' growth community in order to check for electronic uniformity at domain boundaries of synthesized 2D materials and heterostructures over large areas alongside morphological uniformity.

*Methods*

*Sample preparation:* The method of sample preparation is the same as in our previous publications.[33,69] EG is grown at 1800 ºC in a three-phase, hot-zone, graphite furnace *via* silicon sublimation from the 6H SiC(0001) face. It is then patterned using standard ultraviolet photolithography, and a mixture of oxygen and argon (O$_2$/Ar) is used for a reactive ion etch to remove the EG outside of the patterns, leaving behind periodically spaced graphene rectangles of fixed length (5 µm), that ultimately constitute the contacts to the MoS$_2$ channel. Powder vaporization in a horizontal quartz tube furnace is used to synthesize MoS$_2$ at 800 ºC just outside the graphene rectangles, using 2-3 mg of molybdenum trioxide (MoO$_3$) powder and 200 mg sulfur (S) as the precursors. Following the synthesis of EG/MoS$_2$ lateral heterostructure, an SF$_6$/O$_2$ reactive ion etch is used to isolate the MoS$_2$ between the graphene electrodes. Contact regions are then lithographically patterned, briefly exposed to an O$_2$ plasma, and Ti/Au (5/15 nm) metal is deposited *via* electron-beam evaporation, followed by lift off in PRS 3000 photoresist remover. Ti is evaporated onto the sample at a vaccum of ~ 10$^{-9}$ Torr in an electron beam evaporation chamber at a rate of ~ 0.5 Å/s upto a thickness of ~ 5 nm. Au is then evaporated at the rate of ~ 1 Å/s upto a thickness of ~ 20 nm. The sample chuck is cooled to 0 ºC during the entire deposition process.

*Photocurrent measurements:* We measure the photocurrent by scanning a tunable Ar/Kr laser at 488 nm (2.54 eV) across the area of the investigated lateral heterojunctions with piezo stages. The laser is focused on the heterojunctions by a 100× Mitutoyo Plan Apo objective (f = 200 mm) to a diffraction limited spot size of ~ 0.8 µm (FWHM). The resulting photocurrent at each point is measured by a succession of a current preamplifier (Ithaco 1211) at a sensitivity of 10$^{-7}$ A/V and a digital multimeter (Agilent 34401A). All measurements are performed in vacuum (10$^{-6}$ mbar) at room temperature. The SiC substrate does not contribute to the photocurrent at the wavelength of the laser utilized.

*Computational Details:* We utilize the plane-wave Density Functional Theory code QUANTUM-ESPRESSO.[70] We create 4 layer thick slabs of each material as shown in Fig. 2, which was determined to be sufficient to converge the Fermi energy to 50 meV. We utilize Perdew-Burke-Ernzerhof (PBE) exchange-correlation functionals[71] with norm conserving Vanderbilt pseudopotentials from the PseudoDojo library, for all the calculations in the main text.[72] For comparison in Table S.2.1, we also use PBEsol,[73] PZ,[74] and PW91[75] PAW pseudopotentials from the PSLibrary.[76] We sample the Brillioun zone with a 4 x 4 x 1 Monkhorst-Pack grid and 0.0001 Ry of Marzari-Vanderbilt smearing.[77] We select wavefunction and charge density kinetic energy cutoffs of 50 Ry and 200 Ry, respectively. By aligning the potential of the vacuum region to zero, we can obtain the wavefunction of different slabs as the negative of the Fermi level (*i.e.* $\phi = -\epsilon_F$). We follow the procedure outlined in the text to obtain the Schottky barrier of the material, see supplementary section S2 for a summarized procedure. The slabs and interface were generated using the pymatgen utility.[78]

*I-V-T measurements:* Electrostatic double layer (EDL) gating is implemented in an Oxford Optistat closed-cycle helium cooled pulse tube cryostat using $(PEO)_{76}$:$CsClO_4$ as the electrolyte with mobile ions, allowing the I-V-T measurements in the temperature range of 3-300K. Biasing the EG/MoS$_2$/EG and metal/MoS$_2$/EG channel ($V_{ds}$) and biasing the gate ($V_g$) is both performed with a dual-channel Keysight B2912A Precision Source/Measure Unit. In order to arrest the mobile ions in the electrolyte dielectric at a certain gate voltage, the $V_g$ is applied at room temperature and then the system was cooled down to 3K. The temperature is slowly raised in steps of 5K until 300K and source-drain current ($I_{ds}$) was measured at each temperature at a fixed $V_{ds}$ of 100 mV. The $V_g$ is then increased by 0.5V and the measurement cycle is repeated. A side gate geometry is implemented.

*Nano-ARPES:* Angle-resolved photoemission intensity maps were recorded using a focused synchrotron beam and a Scienta R4000 analyzer at the MAESTRO beamline of the Advanced Light Source. A Fresnel zone plate was used to focus the beam. The zone plate used in the experiments allows a minimum spot size of 120 nm. For ARPES, the photon energy was set to Eph = 98 eV, the detector resolution was 125 meV, the entrance slit width and height were 50 μm and the sample was held at room temperature. From the nano-ARPES maps, we determined a spatial resolution of ~300 nm for the measurements presented. The base pressure during the measurements was below $10^{-11}$ mbar. ARPES was conducted on the same samples as discussed in Ref 69.[69] The samples were synthesized *ex situ* (as detailed in Ref 33),[33] exposed to ambient conditions, and then transferred into the nano-ARPES chamber. Prior to the X-ray measurements, the samples were annealed at 100 °C under vacuum for 30 min to remove surface adsorbates.

*Acknowledgments*

S.S. and J.A.R. acknowledge the funding from NSF CAREER (Award: 1453924). S.M. acknowledges support by the Swiss National Science Foundation (Grant No. P300P2-171221). This research used resources of the Advanced Light Source, which is a DOE Office of Science User Facility under Contract No. DE-AC02-05CH11231. P.S. acknowledges financial support from the Alexander von Humboldt foundation and the German federal ministry of education and research within the Feodor Lynen program. D.S. acknowledges funding from the National Science Foundation *via* award number EFMA 1433378 and EFMA 1433307. H.A.S. was supported by the EXPEC Advanced Research Center, Petroleum Engineering and Development Department at Saudi Aramco. D.W. and R.M.F.'s contribution was supported in part by the Center for Low Energy Systems Technology (LEAST), one of six centers of STARnet, a Semiconductor Research Corporation program sponsored by MARCO and DARPA. I.D. acknowledges the financial support from the National Science Foundation under Grant No. DMREF- 1729338. U.W. was funded by the Deutsche Forschungsgemeinschaft (DFG, German Research Foundation) under Germany's Excellence Strategy – EXC 2089/1 – 390776260. This work was performed, in part, at the Center for Integrated Nanotechnologies, an Office of Science User Facility operated for the U.S. Department of Energy (DOE) Office of Science. Sandia National Laboratories is a multimission laboratory managed and operated by National Technology & Engineering Solutions of Sandia, LLC, a wholly owned subsidiary of Honeywell International, Inc., for the U.S. DOE's National Nuclear Security Administration under contract DE-NA-0003525. The views expressed in the article do not necessarily represent the views of the U.S. DOE or the United States Government.
*References*

(1) Liu, X.; Galfsky, T.; Sun, Z.; Xia, F.; Lin, E.; Lee, Y.-H.; Kéna-Cohen, S.; Menon, V. M. Strong Light–Matter Coupling in Two-Dimensional Atomic Crystals. *Nat. Photonics* **2015**, *9* (1), 30–34.

(2) Wu, S.; Buckley, S.; Jones, A. M.; Ross, J. S.; Ghimire, N. J.; Yan, J.; Mandrus, D. G.; Yao, W.; Hatami, F.; Vučković, J.; Majumdar, A.; Xu, X. Control of Two-Dimensional Excitonic Light Emission via Photonic Crystal. *2D Mater.* **2014**, *1* (1), 011001.

(3) Das, S.; Gulotty, R.; Sumant, A. V.; Roelofs, A. All Two-Dimensional, Flexible, Transparent, and Thinnest Thin Film Transistor. *Nano Lett.* **2014**, *14* (5), 2861–2866.

(4) Kim, S. J.; Choi, K.; Lee, B.; Kim, Y.; Hong, B. H. Materials for Flexible, Stretchable Electronics: Graphene and 2D Materials. *Annu. Rev. Mater. Res.* **2015**, *45* (1), 63–84.

(5) Briggs, N.; Subramanian, S.; Lin, Z.; Li, X.; Zhang, X.; Zhang, K.; Xiao, K.; Geohegan, D.; Wallace, R.; Chen, L.-Q.; Terrones, M.; Ebrahimi, A.; Das, S.; Redwing, J.; Hinkle, C.; Momeni, K.; van Duin, A.; Crespi, V.; Kar, S.; Robinson, J. A. A Roadmap for Electronic Grade 2D Materials. *2D Mater.* **2019**, *6* (2), 022001.

(6) Zhang, W.; Wang, Q.; Chen, Y.; Wang, Z.; Wee, A. T. S. Van Der Waals Stacked 2D Layered Materials for Optoelectronics. *2D Mater.* **2016**, *3* (2), 022001.

(7) Wang, Q. H.; Kalantar-Zadeh, K.; Kis, A.; Coleman, J. N.; Strano, M. S. Electronics and Optoelectronics of Two-Dimensional Transition Metal Dichalcogenides. *Nat. Nanotechnol.* **2012**, *7* (11), 699–712.

(8) Mak, K. F.; Shan, J. Photonics and Optoelectronics of 2D Semiconductor Transition Metal Dichalcogenides. *Nat. Photonics* **2016**, *10* (4), 216–226.

(9) Parzinger, E.; Hetzl, M.; Wurstbauer, U.; Holleitner, A. W. Contact Morphology and Revisited Photocurrent Dynamics in Monolayer MoS2. *npj 2D Mater. Appl.* **2017**, *1* (1), 40.

(10) Mak, K. F.; McGill, K. L.; Park, J.; McEuen, P. L. Valleytronics. The Valley Hall Effect in MoS$_2$ Transistors. *Science* **2014**, *344* (6191), 1489–1492.

(11) Deng, D.; Novoselov, K. S.; Fu, Q.; Zheng, N.; Tian, Z.; Bao, X. Catalysis with Two-Dimensional Materials and Their Heterostructures. *Nat. Nanotechnol.* **2016**, *11* (3), 218–230.

(12) Li, H.; Xiao, J.; Fu, Q.; Bao, X. Confined Catalysis under Two-Dimensional Materials. *Proc. Natl. Acad. Sci.* **2017**, *114* (23), 5930–5934.

(13) Machado, B. F.; Serp, P. Graphene-Based Materials for Catalysis. *Catal. Sci. Technol.* **2012**, *2* (1),
12

# Supplementary Information:

# Photo-physics and electronic structure of lateral graphene/MoS$_2$ and metal/MoS$_2$ junctions

Shruti Subramanian[1,2], Quinn T. Campbell[1,3], Simon Moser[4,5], Jonas Kiemle[6], Phillipp Zimmermann[6], Paul Seifert[6,7], Florian Sigger[6], Deeksha Sharma[8], Hala Al-Sadeg[1], Michael Labella III[9], Dacen Waters[10], Randall M. Feenstra[10], Roland J. Koch[4], Chris Jozwiak[4], Aaron Bostwick[4], Eli Rotenberg[4], Ismaila Dabo[1], Alexander Holleitner[6], Thomas E. Beechem[11], Ursula Wurstbauer[6,12], Joshua A. Robinson[1,2,13,14]

1. Department of Materials Science and Engineering, The Pennsylvania State University, University Park, Pennsylvania 16802, United States of America
2. Center for 2-Dimensional and Layered Materials, The Pennsylvania State University, University Park, Pennsylvania 16802, United States of America
3. Center for Computing Research, Sandia National Laboratories, Albuquerque, New Mexico 87185, United States of America
4. Advanced Light Source, E. O. Lawrence Berkeley National Laboratory, Berkeley, California 94720, United States of America
5. Physikalisches Institut and Würzburg-Dresden Cluster of Excellence ct.qmat, Universität Würzburg, 97074 Würzburg, Germany
6. Walter Schottky Institut and Physik Department, Technische Universität München, Am Coulombwall 4, 85748 Garching, Germany
7. ICFO - Institut de Ciencies Fotoniques, The Barcelona Institute of Science and Technology, Castelldefels, Barcelona, 08860, Spain
8. Department of Mechanical Engineering, The Pennsylvania State University, University Park, Pennsylvania 16802, United States of America
9. Nanofabrication Facility, The Pennsylvania State University, University Park, Pennsylvania 16802, United States of America
10. Department of Physics, Carnegie Mellon University, Pittsburgh, PA 15213, United States of America
11. Center for Integrated Nanotechnologies, Sandia National Laboratories, Albuquerque, New Mexico 87185, United States of America
12. Institute of Physics, University of Munster, 48149 Münster, Germany
13. 2-Dimensional Crystal Consortium, The Pennsylvania State University, University Park, PA 16802, United States of America
14. Center for Atomically Thin Multifunctional Coatings, The Pennsylvania State University, University Park, PA 16802, United States of America

**Supplementary Section S.1:**

The asymmetric heterostructure shown in Figure 1(b) and 1(d) of the main manuscript demonstrates a less uniform photocurrent at the heterostructure interfaces when compared to the symmetric heterostructure shown in Figure 1(a) and 1(c) of the main manuscript. We take line cuts at 4-5 different y- values for both the symmetric and asymmetric heterostructure in Figure S.1.1 to demonstrate the uniformity of the former when compared to the latter. The linecuts in the symmetric heterostructure of Figure S.1.1(a) have very similar photocurrent landscapes as shown in Figure S.1.1(b), demonstrating the uniformity of the built-in field through the length of the measured device at the photo-active heterostructure interfaces. In contrast, linecuts in the asymmetric heterostructure of Figure S.1.1(c) have large variations in the measured photocurrent landscapes as shown in Figure S.1.1(d), demonstrating the non-uniformity of the built-in field through the length of the measured device. Figure S.1.1(d) is a demonstration of the possible formation of hot spots due to current crowding. A >10× increase in the measured photocurrent in the symmetric device compared to the asymmetric one is shown in Figure S.1.2.

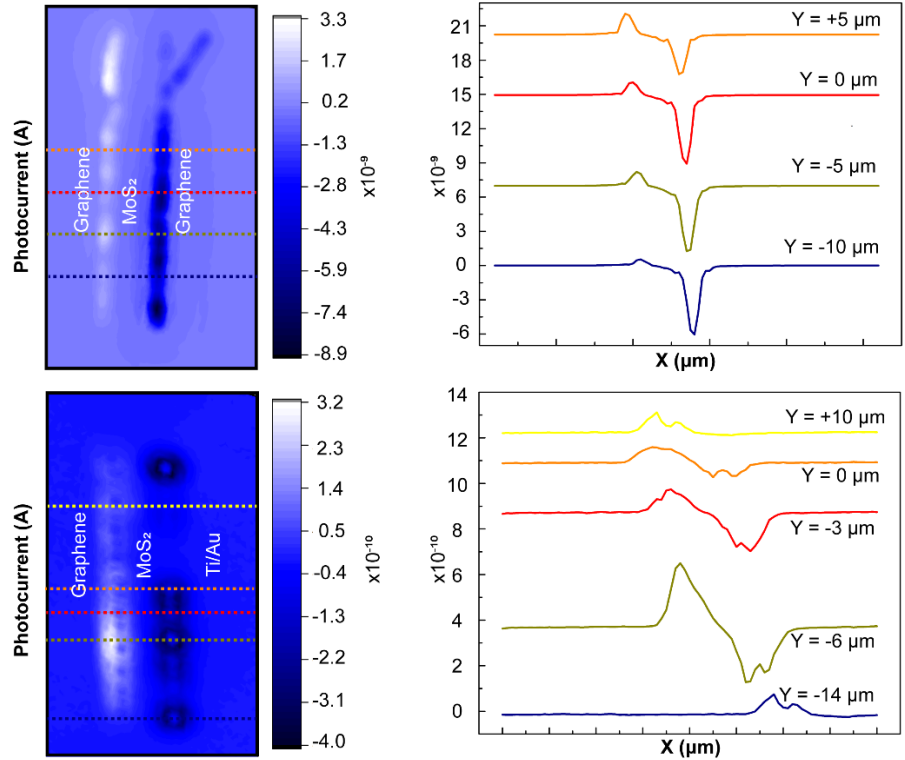

**Figure S.1.1:** (a) Photocurrent map (at zero applied bias) of the EG/MoS$_2$/EG heterostructure displaying photo-activity at the EG/MoS$_2$ interface. (b) Linecuts from (a) show a uniform photocurrent landscape through the length of the symmetric heterostructure device. (c) Photocurrent map (at zero applied bias) of the metal/MoS$_2$/EG heterostructure displaying "patchy" photo-activity at the metal/MoS$_2$ interface as compared to the EG/MoS$_2$ interface. (d) Linecuts from (c) show non-uniform photocurrent landscape through the length of the asymmetric heterostructure device.

Note that the magnitude of photocurrent is not the same at the two EG/MoS$_2$ heterojunctions of the symmetric heterostructure as shown in the photocurrent landscapes in Fig S.1.1(b). This difference in magnitude of photocurrent can be attributed to local inhomogeneities at the EG/MoS$_2$ heterostructure interface caused by variations due to powder vaporization synthesis, and lithographic variations with different lengths of exposed graphene on both sides, thus contributing different resistances to the motion of electrons and holes at the heterostructure interface. In an ideal scenario with a perfectly symmetrical EG/MoS$_2$/EG device, the magnitude of photocurrent should be equal on both sides.

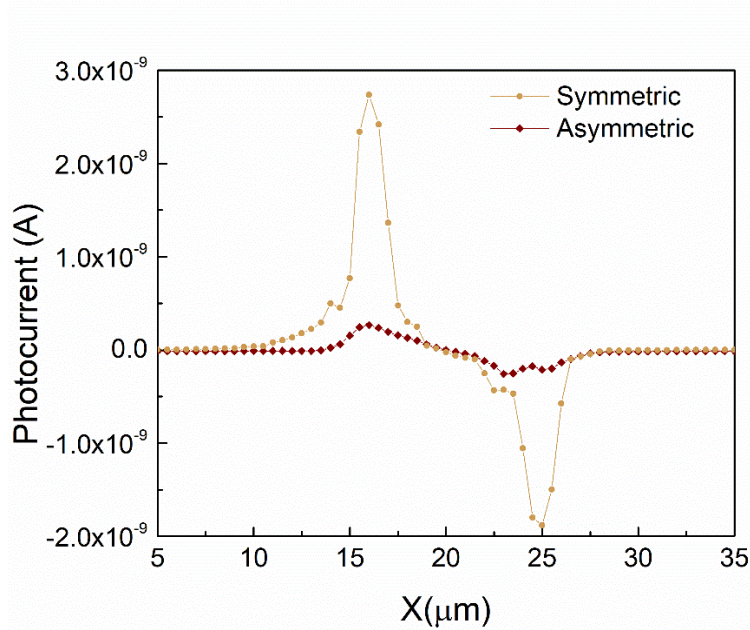

**Figure S.1.2:** Linecuts from photocurrent maps comparing the symmetric and asymmetric heterostructure, showing the >10× increase in photocurrent in the symmetric heterostructure as compared to the asymmetric heterostructure.

We perform statistical analysis on the photocurrent obtained at the two interfaces of the asymmetric heterostructure to demonstrate the non-uniformity of the metal/MoS$_2$ interface. The standard deviation (σ) values for the EG/MoS$_2$ interface in the asymmetric heterostructure is $\sim 6 \times 10^{-11}$ A and that for the metal/MoS$_2$ interface is $\sim 9 \times 10^{-11}$ A, making the metal/MoS$_2$ 1.5× less uniform as compared to the EG/MoS$_2$ in the asymmetric heterostructure. This indicates that the metal/MoS$_2$ heterointerface has more undulations through the length of the device as compared to the EG/MoS$_2$ interface, directly corresponding to the lack of consistent photocurrent signal through the length of the device as seen in Figure 1(f) of the main manuscript.

**Supplementary Section S.2:**

The total voltage drop ($\Delta\phi$) across a junction is the difference between the work functions of the isolated interface and the metal contact. At an ideal metal/semiconductor interface, the entire $\Delta\phi$ is accommodated by the bulk semiconductor. In contrast, if there is 100% Fermi level pinning, all the $\Delta\phi$ is captured by interfacial trap making the Schottky barrier a constant value independent of the metal work function. A realistic interface exhibits a band bending profile between these two extremes, with some amount of $\Delta\phi$ accommodated by the interfacial trap states (the interfacial barrier), and the remaining $\Delta\phi$ accommodated by the depletion layer of the semiconductor (the Schottky barrier).

We utilize a previously developed technique to quantify the $\Delta\phi$ accommodated by the Schottky barrier and the interfacial charges.[1,2] A limited $\Delta\phi$ is introduced across the metal/MoS$_2$ and EG/MoS$_2$ interface within DFT. The fraction of this $\Delta\phi$ compensated by the interfacial charges is measured and the remaining $\Delta\phi$ results in the slope of the potential in the semiconductor depletion layer. The Schottky barrier is then extrapolated for this particular $\Delta\phi$ using the Mott Schottky equations, which are derived by using the Poisson equation and assuming Boltzmann statistics for the charge within the semiconductor (see Schmickler and Santos Ch. 11 for a full derivation).[3] Equation 1 of the main text is the final result of this exercise.

Introducing several $\Delta\phi$ values generates a wide range of Schottky barriers, but the interface will only be at equilibrium with the doped EG or the Ti for one value of $\Delta\phi$. The equilibrium Schottky barrier is determined by testing a wide range of $\Delta\phi$ values across the interface until we find the $\Delta\phi$ that leads to the work function of the interface and the work function of doped EG being equal. The cutoff plane is placed two layers within the semiconductor, allowing the interface potential to be measured while avoiding spurious surface effects from the other end of the semiconductor slab. Following the determination of the equilibrium $\Delta\phi$, we measure the slope of the potential at the cutoff plane and extrapolate the Schottky barrier using Equation 1.

Finally, our DFT calculations support the hypothesis that the EG/MoS$_2$ interface bond is van der Waals in character and the Ti/MoS$_2$ interface bond is more covalent/ionic. We perform Bader charge analysis[4] on both neutral interfaces to recover the amount of electrons that can be attributed to each atom. In the EG/MoS$_2$ interface, sulfur atoms have an average of 6.6 $e$ per atom, which remains constant both at the interface and in the bulk MoS$_2$ section. In contrast, at the Ti/MoS$_2$ interface, sulfur atoms have an average of 7.4 $e$ at the interface, and a more typical average of 6.8 $e$ in the bulk MoS$_2$ section. This indicates that there is a significant amount of charge trapping at the Ti/MoS$_2$ interface that does not take place at the EG/MoS$_2$ interface, suggesting a more covalent or ionic character to the bonds at the Ti/MoS$_2$ interface, while the EG/MoS$_2$ bonds do not involve charge trapping and are more likely to be van der Waals in character.

**Table S.2.1:** The work functions of both isolated materials and interfaces as calculated from DFT using multiple functionals.

| Material | Work function PBE (eV) | Work function PBEsol (eV) | Work function PZ (eV) | Work function PW91 (eV) |
|---|---|---|---|---|
| Pristine graphene ($\phi_{gr}$) | 3.92 | 3.95 | 4.17 | 3.99 |
| MoS$_2$ ($\phi_{MoS_2}$) | 3.78 | 3.77 | 4.00 | 3.79 |
| SiC ($\phi_{SiC}$) | 4.06 | 4.02 | 5.18 | 4.05 |
| Ti ($\phi_{Ti}$) | 4.02 | 4.03 | 4.05 | 4.07 |
| EG/MoS$_2$ interface ($\phi_{EG/MoS_2}$) | 3.55 | 3.65 | 3.88 | 3.62 |
| Ti/MoS$_2$ interface ($\phi_{Ti/MoS_2}$) | 3.21 | 2.88 | 3.04 | 2.85 |

**Supplementary Section S.3:**

The Arrhenius equation is used to extract Schottky barrier heights ($\phi_s$) from temperature dependent current (I-V-T) measurements as explained by Das et al.[5] Two types of devices are used – MoS$_2$ contacted by EG and MoS$_2$ contacted by metal (Ti/Au). Electric double layer (EDL) gating is employed in a helium-cooled cryostat and the temperature range used is 3-300K and the gate voltage ($V_g$) range used is -2.5V to +2.5V. However, for equation (i) below to be valid, it is essential to be in the thermionic region of the transistor operation. Hence, the actual data used for Schottky barrier extraction in Figure S.3.1 is between 150-300K, because all temperatures below this range do not exhibit a thermal excitation of carriers over the Schottky barrier. The $V_g$ range shown in Figure S.3.1 eliminates the noise floor in the *off* region of transistor operation and also eliminates the tunneling region of the *on* state of the transistor. The device characteristics shown have been obtained *via* a side gated geometry and application of a 100 mV source-drain bias ($V_{ds}$).

$$I_{ds} = AT^{1.5} \exp\left(\frac{-q\phi_B}{kT}\right)\left(1 - \exp\left(\frac{-qV_{ds}}{\eta kT}\right)\right) \qquad \text{Equation (i)}$$

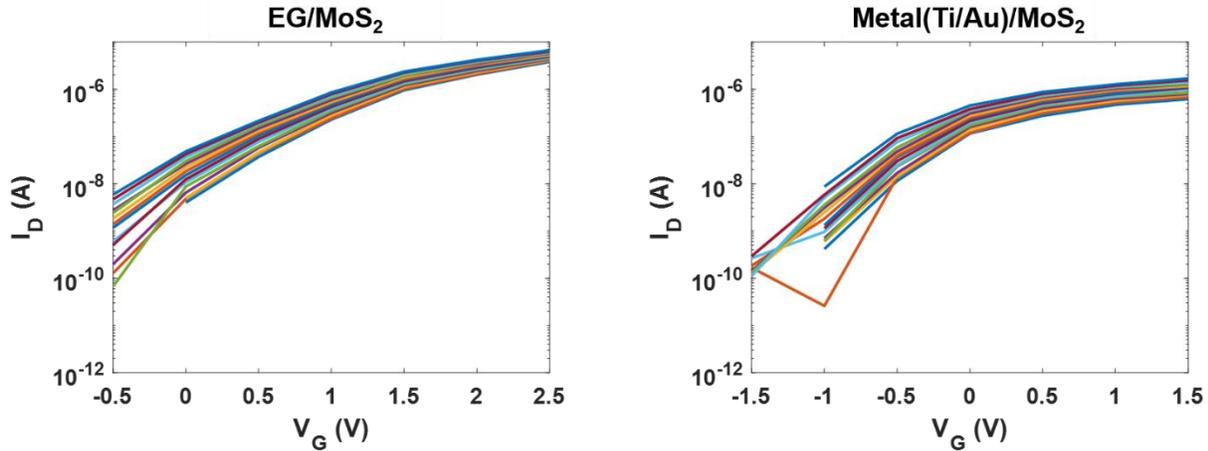

**Figure S.3.1:** Transfer curves ($I_{ds}$ vs. $V_g$) obtained for EG/MoS$_2$ and metal/MoS$_2$ using a 100 mV $V_{ds}$ and electric double layer side gate geometry for a temperature range of 150-300K.

The second exponential term $\exp\left(\frac{-qV_{ds}}{\eta kT}\right)$ is small and independent of the value of the ideality factor $1 < \eta < 2$, making the term $\left(1 - \exp\left(\frac{-qV_{ds}}{\eta kT}\right)\right) \sim 1$ for all cases. Equation (i) is simplified to look like equation (ii) below. The transfer curves from Figure S.3.1 are rearranged to look like Figure S.3.2 where $\ln\left(\frac{I_{ds}}{T^{1.5}}\right)$ i.e. the left hand side of equation (ii) is plotted against $1/T$ i.e. the right hand side of the variable T, for different $V_g$ values.

$$\ln\left(\frac{I_{ds}}{T^{1.5}}\right) = -\frac{q\phi_B}{k_B T} + \ln(A) \qquad \text{Equation (ii)}$$

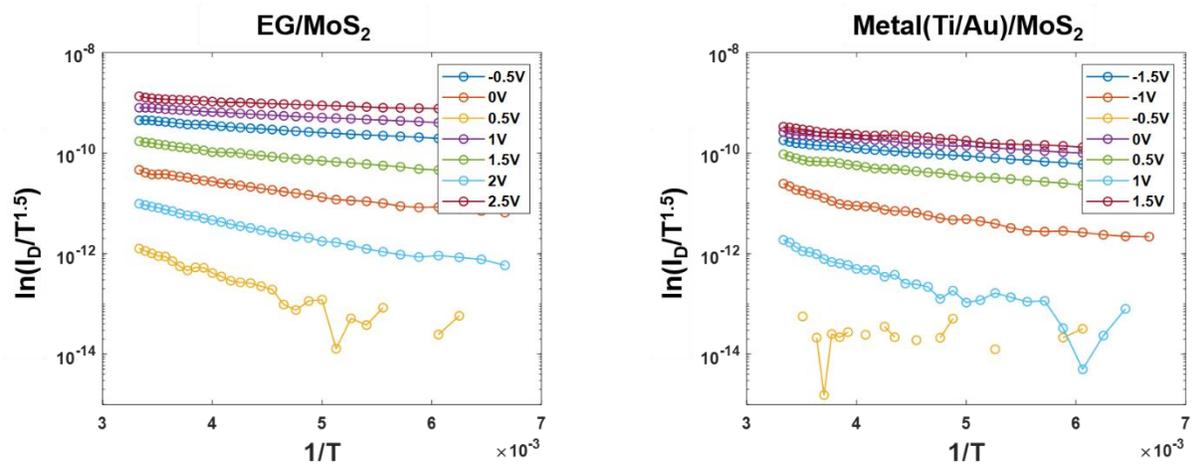

**Figure S.3.2:** According to the Arrhenius equation rearrangement in equation (ii), the transfer curves from Figure S.3.1 are replotted as $ln\left(\frac{I_{ds}}{T^{1.5}}\right)$ vs. $1/T$.

The slope of the curves in Figure S.3.2 is then used to extract the Schottky barrier heights ($\phi_B$) that are shown in Figure 2(e). For all values of $V_g$, the *apparent* Schottky barrier is larger for the metal/MoS$_2$ as compared to the EG/MoS$_2$.

**Supplementary Section S.4:**

The spatial variations in the electronic band structure at the synthesized EG/MoS$_2$ lateral heterostructure interface are investigated at steady state using a synchrotron-based angle resolved photoemission spectroscopy setup with a nanometer scale resolution (nano-ARPES). The experimental setup to obtain the 4D dataset *via* nano-ARPES is shown in Figure S.5.1. Room temperature nano-ARPES results are obtained with 98 eV photons and an X-ray spot size of ~300 nm FWHM. The X-ray probing spot is focused using a Fresnel zone plate and scanned across the sample. The ARPES band structure I(E,k) is measured at every spatial point coordinate (x,y), E is the binding energy referenced to the Fermi level and k is the in-plane momentum along the orientation of the analyzer entrance slit as shown in Figure S.4.1. In this scanning mode, nano-ARPES is particularly suited for flat conductive samples like ours, providing a well-defined surface normal, and thus a well-preserved momentum resolution. The nano-ARPES results are detailed in Figure 3 of the main manuscript.

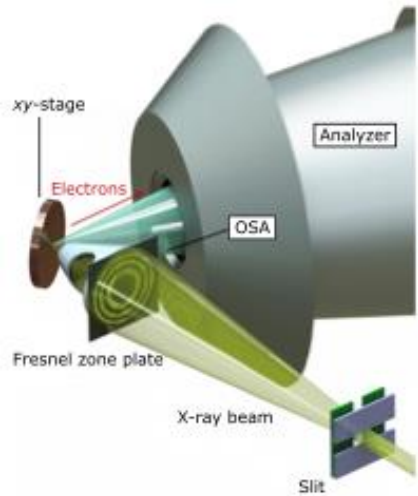

**Figure S.4.1:** Schematic of the nanoscale photoemission spectroscopy setup of the MAESTRO beamline at the Advanced Light Source. The X-ray beam from the entrance slit is imaged with the help of a Fresnel zone plate and an order sorting aperture (OSA) onto the sample (minimum spot size ~120 nm). Locally emitted photoelectrons are collected by the analyzer, and spatial resolution is achieved by scanning the sample stage.

A darker contrast in the nano-ARPES maps indicates a higher intensity. The scale bars of the various nano-ARPES maps of Figure 3 in the main manuscript are shown in Figure S.4.2 below. The *max* and *min* indicates the higher and lower intensity of ARPES signal, respectively.

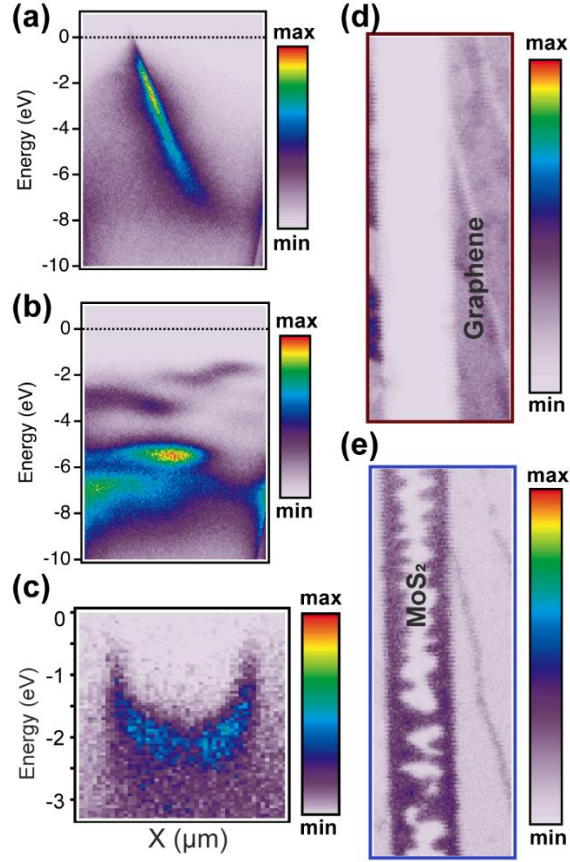

**Figure S.4.2:** Nano-ARPES showing the scale bars to explain the various color intensities shown in Figure 3 of the main manuscript. Due to slight variations in contrast used for all panels, their individual scale bars have been shown (a) shows the Dirac band of graphene (same as Figure 3(b) from the main manuscript). (b) shows the valence band states of $MoS_2$ (same as Figure 3(c) from the main manuscript). (c) Bending of the valence band maximum of the $MoS_2$ (same as Figure 3(g) and 3(h) from the main manuscript). (d) and (e) show the spatial distribution of graphene and $MoS_2$ through the sample with their intensity scale bars (same as Figure 3(b1) and 3(c1) from the main manuscript).

In the nano-ARPES experiment, the entrance slit of the electron analyzer is aligned perpendicular to the channel, but *not* perfectly aligned with respect to particular high symmetry directions of graphene or $MoS_2$. As a careful band mapping could not be done, we cannot annotate the photoelectron emission angles precisely to their corresponding momenta. From their band dispersions, however, we can correlatively associate regions as "close to the Γ point" and "close to the K" point for the $MoS_2$ and graphene respectively. A precise annotation would be misleading, however.

ARPES is not a suitable probe to precisely determine the electronic work function. ARPES primarily measures the kinetic energy spectrum, *i.e.* $E_{kin} = h\nu - E_b - \phi$, where the work function $\phi$ corresponds to the work function of the photoelectron analyzer, and is not influenced by the work function of the sample (which is a result of the contact potential between sample and analyzer that "automatically compensates" for any change of the substrate work function). In principle, the work function of the sample can be determined from the cutoff energy of secondary electrons. But as this determination of the cutoff is not precise and the corresponding experiment is very time consuming in the limited amount of available machine time, it was omitted.

Further details from the nano-ARPES measurements as seen in Figure 3 of the main manuscript are encompassed in the figures below.

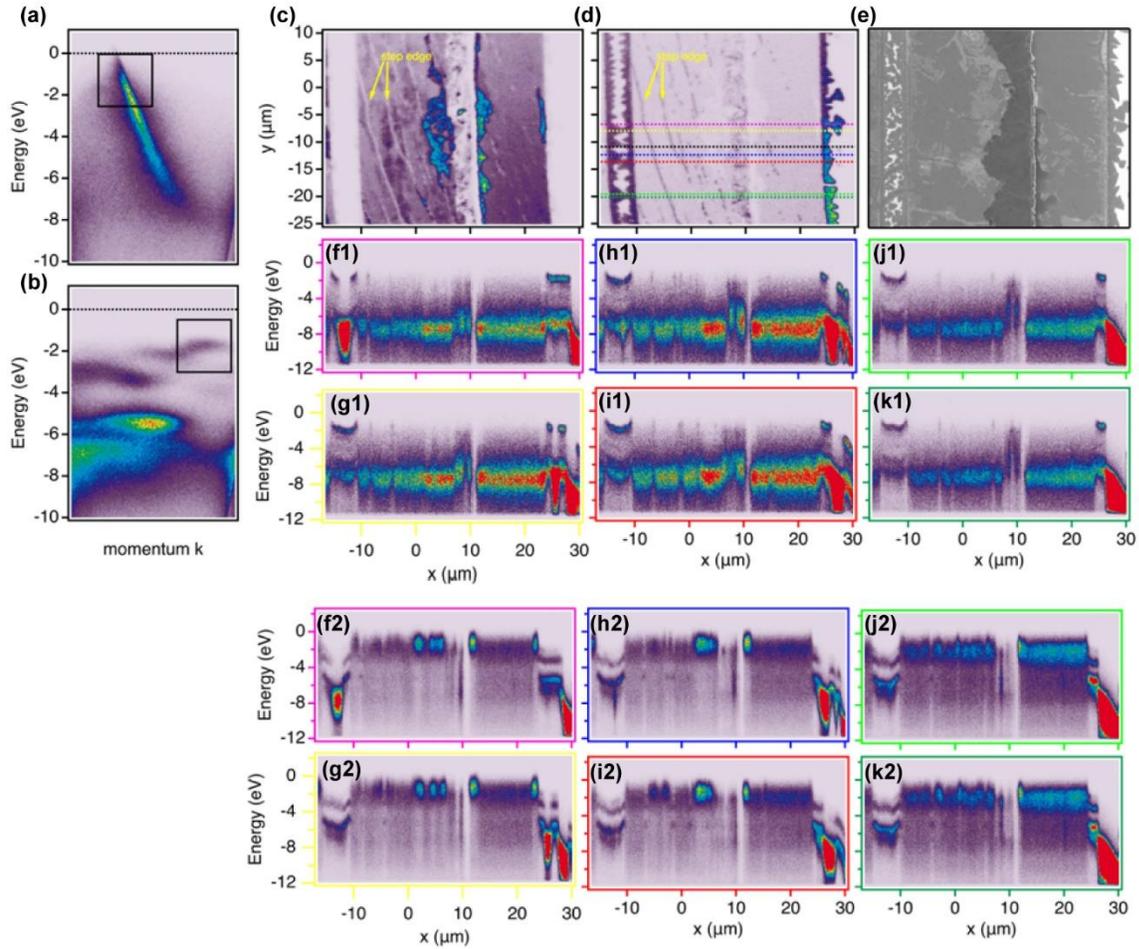

**Figure S.4.3:** Nano-ARPES across the entire lateral EG/MoS$_2$ heterostructure. (a) ARPES band structure isolating the π bands of graphene. (b) Nano-ARPES band structure isolating the valence band of MoS$_2$. (c) Intensity integrated over the (E,k) window in (a), showing the distribution of the graphene as a function of (x,y). (d) Intensity integrated over the (E,k) window in (b), showing the distribution of the MoS$_2$ as a function of (x,y). (e) SEM image across a similar channel. (f-k) E *vs.* x ARPES cuts extracted from (d), showing (f1-k1) the band bending of the MoS$_2$ valence band and the variation of the graphene pi-band (f2-k2) across the entire sample.

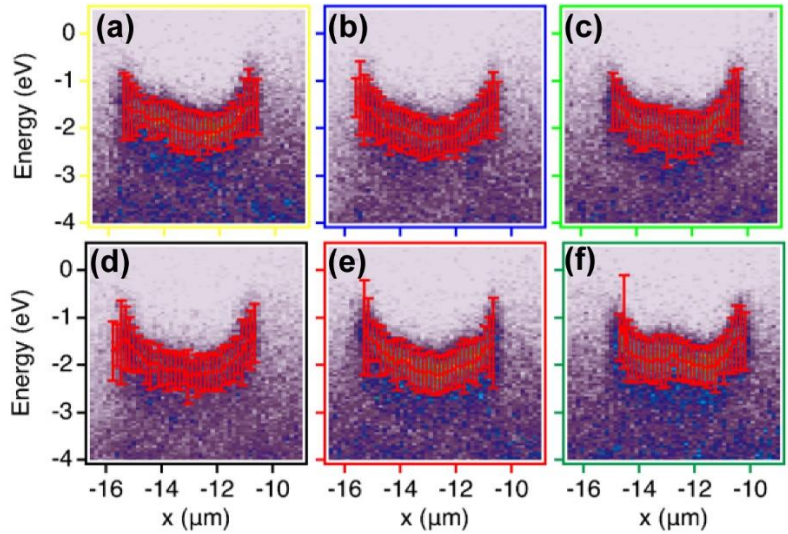

**Figure S.4.4:** E *vs.* x ARPES cuts across the MoS$_2$ channel extracted along the colored horizontal lines in Figure S.4.2(d) and corresponding to the colored panels of Fig. S.4.2.(f-k). The MoS$_2$ valence band onset is fitted by a Gaussian peak whose position and width is shown by red markers and error bars, respectively.

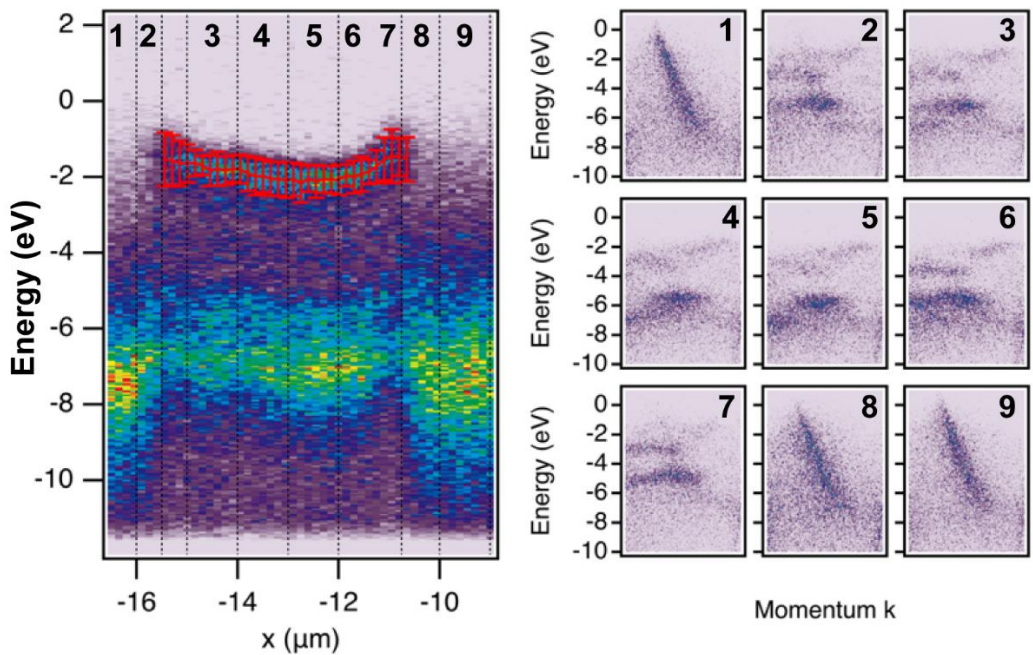

**Figure S.4.5:** E *vs.* x ARPES cut across the MoS$_2$ channel extracted along the black horizontal line in Figure S.4.3(d). (1-9) Individual E *vs. k* ARPES maps collected at selected positions x as indicated in figure on the left.